\documentclass[twocolumn,aps,pra,showpacs]{revtex4}
\usepackage[T1]{fontenc}
\usepackage[latin9]{inputenc}
\usepackage{amsmath}
\usepackage{amssymb}
\usepackage{graphicx}
\usepackage{esint}

\makeatletter
\@ifundefined{textcolor}{}
{%
 \definecolor{BLACK}{gray}{0}
 \definecolor{WHITE}{gray}{1}
 \definecolor{RED}{rgb}{1,0,0}
 \definecolor{GREEN}{rgb}{0,1,0}
 \definecolor{BLUE}{rgb}{0,0,1}
 \definecolor{CYAN}{cmyk}{1,0,0,0}
 \definecolor{MAGENTA}{cmyk}{0,1,0,0}
 \definecolor{YELLOW}{cmyk}{0,0,1,0}
}

\makeatother

\makeatother

\begin{document}

\title{First and second sound of a unitary Fermi gas in highly elongated
harmonic traps}

\author{Xia-Ji Liu$^{1}$ and Hui Hu$^{1}$}

\affiliation{$^{1}$Centre for Quantum and Optical Science, Swinburne University
of Technology, Melbourne 3122, Australia}

\date{\today}
\begin{abstract}
Using a variational approach, we present the full solutions of the
simplified one-dimensional two-fluid hydrodynamic equations for a
unitary Fermi gas trapped in a highly elongated harmonic potential,
which is recently derived by Stringari and co-workers {[}Phys. Rev.
Lett. \textbf{105}, 150402 (2010){]}. We calculate the discretized
mode frequencies of first and second sound along the weak axial trapping
potential, as a function of temperature and the form of superfluid
density. We show that the density fluctuations in second sound modes,
due to their coupling to first sound modes, are large enough to be
measured in current experimental setups such as that exploited by
Tey \textit{et al}. at the University of Innsbruck {[}Phys. Rev. Lett.
\textbf{110}, 055303 (2013){]}. Owing to the sensitivity of second
sounds on the form of superfluid density, the high precision of the
measured second sound frequencies may provide us a promising way to
accurately determine the superfluid density of a unitary Fermi gas,
which so far remains elusive.
\end{abstract}


\maketitle

\section{Introduction}

Over the past few years, first and second sound of a unitary atomic
Fermi gas at a broad Feshbach resonance have received increasing attentions
\cite{Taylor2005,He2007,Taylor2008,Taylor2009,Watanabe2010,Hu2010,Salasnich2010,Bertaina2010,Sidorenkov2013,Tey2013,Guajardo2013,Hou2013,Hu2013}.
Being the in-phase density oscillations (first sound) and out-of-phase
temperature oscillations (second sound) \cite{Tisza1938,Landau1941},
these sound modes provide a useful probe of the equation of state
\cite{Tey2013,Guajardo2013,Stringari2004,Hu2004,Altmeyer2007,Joseph2007}
and superfluid density \cite{Taylor2008,Taylor2009,Watanabe2010,Hu2010,Salasnich2010,Sidorenkov2013,Hu2013}
of the unitary Fermi gas. The latter quantity of superfluid density
is of particular interest, as it is notoriously difficult to calculate
in theory. In practice, it could be measured through second sound
wave propagation \cite{Watanabe2010}. Indeed, in superfluid helium
$^{4}$He, the accurate knowledge of its superfluid density, slightly
below the lambda transition, is obtained by the measurement of temperature
waves \cite{Dash1957}.

A strongly interacting unitary Fermi gas bears a lot of similarity
to superfluid helium \cite{Taylor2009,Hu2010}. Due to strong correlation,
the first and second sound in both systems are nearly decoupled. Yet,
the weak coupling between sounds still leads to a sizable hybridization
effect and hence a measurable density fluctuation for second sound.
This gives rise to a promising way of exciting and detecting second
sound through density measurements in a unitary Fermi gas \cite{Watanabe2010,Hu2010}.
For an isotropically trapped unitary Fermi gas, such a hybridization
effect has been analyzed by Taylor \textit{et al.}, using the standard
dissipationless Landau two-fluid hydrodynamic approach \cite{Taylor2009}.
Experimentally, however, it is more feasible to confine a unitary
Fermi gas in highly elongated traps. For this configuration, the viscosity
and thermal conductivity terms in the two-fluid hydrodynamic equations
become important and enable a simplified one-dimensional (1D) hydrodynamic
description, as suggested by Bertaina, Pitaevskii and Stringari \cite{Bertaina2010}.
In a recent milestone experiment, a second sound wave has been excited
in a highly elongated unitary Fermi gas, and its propagation along
the weakly confined axial axis has been measured \cite{Sidorenkov2013}.
The simplified 1D hydrodynamic equations has been used to extract
the superfluid density from the resulting second sound velocity data.
Unfortunately, at present the experimental accuracy of sound velocity
is not enough to give a satisfactory determination of superfluid density
\cite{Sidorenkov2013}.

In this paper, we propose that the measurement of \emph{discretized}
mode frequencies of low-lying second sound along the weakly confined
axial direction may provide an accurate means of determining superfluid
density. Indeed, the latest measurement of discretized first sound
mode frequencies \cite{Tey2013} indicates a very small relative error
($\sim0.5\%$), which is at least an order smaller in magnitude than
the relative error in sound velocity data \cite{Sidorenkov2013}.

For this purpose, we fully solve the coupled 1D hydrodynamic equations
in the presence of a weak axial harmonic potential, and obtain the
density fluctuations of discretized low-lying second sound modes,
which arise from to their coupling to first sound modes. We find that
these density fluctuations are significant, thereby making second
sound modes observable in current experiments, by modulating, for
example, the weakly confined axial trapping potential. Our full solutions
of the simplified 1D hydrodynamic equations complement the earlier
results obtained by Hou, Pitaevskii and Stringari \cite{Hou2013},
where the decoupled first and second sound mode frequencies are calculated
with simple variational ansatz for displacement fields. In this work,
we emphasize the correction to discretized mode frequencies, due to
the weak coupling between first and second sound.

The rest of paper is organized as follows. In the next section, we
briefly outline the reduced 1D thermodynamics, as an input for the
simplified 1D hydrodynamic description. In Sec. III, we show how to
solve the coupled 1D hydrodynamic equations by using a variational
approach. In Sec. IV, we first provide results for the decoupled first
and second sound solutions and then present the density fluctuations
of some low-lying second sound modes. Finally, in Sec. V we draw our
conclusion and briefly describe how to obtain the superfluid density
of a unitary Fermi gas from the measured low-lying second sound mode
frequencies.

\section{1D reduced thermodynamics}

We consider a unitary Fermi gas trapped in highly anisotropic harmonic
potential,
\begin{equation}
V_{ext}\left(r_{\perp},z\right)=\frac{1}{2}m\omega_{\perp}^{2}r_{\perp}^{2}+\frac{1}{2}m\omega_{z}^{2}z^{2},
\end{equation}
with atomic mass $m$ and trapping frequency $\omega_{z}\ll\omega_{\perp}$.
We assume that the number of atoms in the Fermi cloud, typically of
$N\sim10^{5}$ in current experiments, is large enough, so that we
can safely use the local density approximation and treat the atoms
in the position $(r_{\perp},z)$ as uniform matter with a local chemical
potential $\mu(r_{\perp},z)=\mu-V_{ext}(r_{\perp},z)$, where $\mu$
is the chemical potential at the trap center. In this way, we may
write the local pressure and number density in the form,
\begin{align}
P\left(r_{\perp},z\right) & =\frac{k_{B}T}{\lambda_{T}^{3}}f_{p}^{3D}\left[\frac{\mu(r_{\perp},z)}{k_{B}T}\right],\\
n\left(r_{\perp},z\right) & =\frac{1}{\lambda_{T}^{3}}f_{n}^{3D}\left[\frac{\mu(r_{\perp},z)}{k_{B}T}\right],
\end{align}
where $\lambda_{T}\equiv\sqrt{2\pi\hbar^{2}/(mk_{B}T)}$ is the thermal
wavelength at temperature $T$, $f_{p}^{3D}\left(t\right)$ and $f_{n}^{3D}\left(t\right)=df_{p}^{3D}\left(t\right)/dt$
are two universal functions satisfied by a unitary Fermi gas due to
its universal thermodynamics \cite{Ho2004,Liu2005,Hu2006a,Hu2006b,Hu2007,Hu2008,Ku2012,Hu2010b}.

It was shown by Bertaina, Pitaevskii and Stringari \cite{Bertaina2010}
that, with tight radial confinement, the standard Landau two-fluid
hydrodynamic equations defined in three dimensions can be greatly
simplified. The key observation is that, as a direct consequence of
the dissipation terms (i.e., nonzero viscosity and thermal conductivity),
the local fluctuations in temperature and chemical potential become
essentially independent on the radial coordinates, if we are interested
in the low-energy excitations at frequency $\omega_{z}\ll\omega_{\perp}$.
Therefore, we could integrate out the radial degree of freedom in
thermodynamic variables and derive 1D reduced thermodynamics \cite{Hou2013}.
In particular, we may obtain a reduced Gibbs-Duhem relation,
\begin{equation}
\delta P_{1}=s_{1}\delta T+n_{1}\delta\mu,
\end{equation}
where the variables $P_{1}$, $s_{1}$ and $n_{1}$ are the radial
integrals of their three-dimensional counterparts, namely the local
pressure, entropy density and number density. For example, we have
\cite{Hou2013},
\begin{equation}
P_{1}\left(z\right)\equiv\int dr_{\perp}2\pi r_{\perp}P\left(r_{\perp},z\right)=\frac{2\pi\left(k_{B}T\right)^{2}}{m\omega_{\perp}^{2}\lambda_{T}^{3}}f_{p}\left(x\right),\label{eq:pressure1d}
\end{equation}
where 
\begin{equation}
x\equiv\left(\mu-\frac{1}{2}m\omega_{z}^{2}z^{2}\right)/k_{B}T
\end{equation}
 and we have introduced the universal scaling function, 
\begin{equation}
f_{p}\left(x\right)\equiv\int_{0}^{\infty}dtf_{p}^{3D}\left(x-t\right).\label{eq:fp}
\end{equation}
 All the 1D thermodynamic variables can then be derived from the reduced
Gibbs-Duhem relation, such as \cite{Hou2013} 
\begin{align}
n_{1}\left(z\right) & =\left(\frac{\partial P_{1}}{\partial\mu}\right)_{T}=\frac{2\pi k_{B}T}{m\omega_{\perp}^{2}\lambda_{T}^{3}}f_{n}\left(x\right),\label{eq:density1d}\\
s_{1}\left(z\right) & =\left(\frac{\partial P_{1}}{\partial T}\right)_{\mu}=\frac{2\pi k_{B}T}{m\omega_{\perp}^{2}\lambda_{T}^{3}}\left[\frac{7}{2}f_{p}\left(x\right)-xf_{n}\left(x\right)\right],\label{eq:entropy1d}
\end{align}
where $f_{n}(x)\equiv df_{p}(x)/dx=f_{p}^{3D}(x)$ according to Eq.
(\ref{eq:fp}). Furthermore, it is straightforward to obtain the specific
heats per particle at constant linear density and pressure \cite{Hou2013},
\begin{align}
\bar{c}_{v1}\left(z\right) & =T\left(\frac{\partial\bar{s}_{1}}{\partial T}\right)_{n_{1}}=\frac{35}{4}\frac{f_{p}\left(x\right)}{f_{n}\left(x\right)}-\frac{25}{4}\frac{f_{n}\left(x\right)}{f_{n}^{'}\left(x\right)},\label{eq:cv1d}\\
\bar{c}_{p1}\left(z\right) & =T\left(\frac{\partial\bar{s}_{1}}{\partial T}\right)_{P_{1}}=\bar{c}_{v1}\frac{7}{5}\frac{f_{p}\left(x\right)f_{n}^{'}\left(x\right)}{f_{n}^{2}\left(x\right)},\label{eq:cp1d}
\end{align}
where $\bar{s}_{1}\equiv s_{1}/(n_{1}k_{B})$ is the entropy per particle
and $f_{n}^{'}(x)\equiv df_{n}(x)/dx=f_{n}^{3D}(x)$. It is also easy
to check the universal relations,
\begin{align}
n_{1}\left(\frac{\partial P_{1}}{\partial n_{1}}\right)_{\bar{s}_{1}} & =\frac{7}{5}P_{1},\label{eq:dpdn}\\
\left(\frac{\partial P_{1}}{\partial s_{1}}\right)_{n_{1}} & =\frac{2}{5}T.\label{eq:dpds}
\end{align}
For the local superfluid density, similarly we express it by a universal
function $f_{s}^{3D}$: 
\begin{equation}
n_{s}\left(r_{\perp},z\right)=\frac{1}{\lambda_{T}^{3}}f_{s}^{3D}\left[\frac{\mu(r_{\perp},z)}{k_{B}T}\right].
\end{equation}
By integrating out the radial coordinate, we find the expression
\begin{equation}
n_{s1}\left(z\right)=\int dr_{\perp}2\pi r_{\perp}n_{s}\left(r_{\perp},z\right)=\frac{2\pi k_{B}T}{m\omega_{\perp}^{2}\lambda_{T}^{3}}f_{s}\left(x\right),\label{eq:sfdensity1d}
\end{equation}
where the universal scaling function $f_{s}(x)$ is defined by, 
\begin{equation}
f_{s}\left(x\right)=\int_{0}^{\infty}dtf_{s}^{3D}\left(x-t\right).\label{eq:fs}
\end{equation}

\begin{figure}
\begin{centering}
\includegraphics[clip,width=0.48\textwidth]{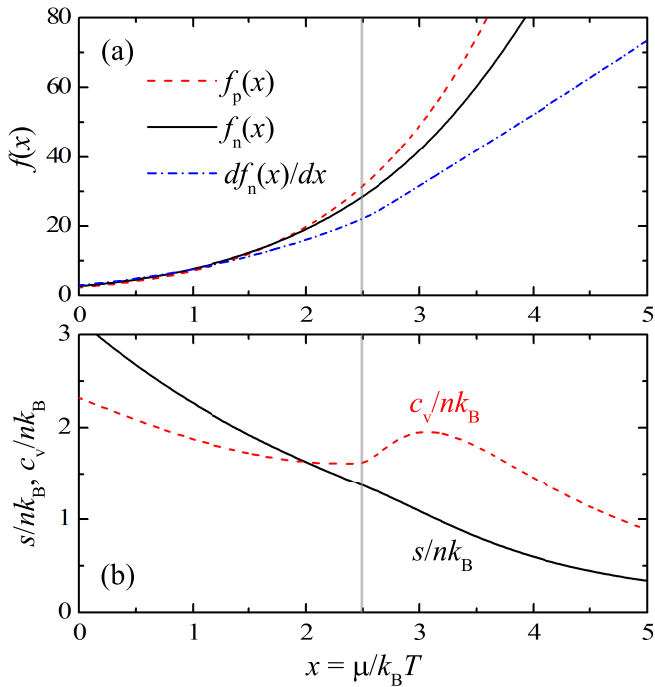} 
\par\end{centering}

\caption{(Color online) (a) 1D universal scaling functions $f_{p}(x)$, $f_{n}(x)$
and $df_{n}(x)/dx$ as a function of the dimensionless variable $x=\mu/(k_{B}T)$.
(b) 1D entropy $\bar{s}=s/(nk_{B})$ and specific heat per particle
$\bar{c}_{v}=c_{v}/(nk_{B})$ as a function of $x=\mu/(k_{B}T)$.
The vertical grey lines indicate the critical threshold for superfluidity,
$x_{c}\simeq2.49$ \cite{Ku2012}.}

\label{fig1} 
\end{figure}

The universal function $f_{p}^{3D}\left(t\right)$ or $f_{n}^{3D}\left(t\right)$
of a homogeneous unitary Fermi gas, where $t\equiv\mu/k_{B}T$, has
been measured by the MIT team with high precision \cite{Ku2012},
both below and above the critical temperature for superfluid phase
transition. In Fig. \ref{fig1}, we show the 1D universal scaling
functions, calculated by using the experimental MIT data for $f_{n}^{3D}\left(t\right)$,
which has been smoothly extrapolated to both low and high temperature
regimes where the behavior of $f_{n}^{3D}\left(t\right)$ is known
\cite{Hu2010,Hou2013,Liu2009,Liu2013}. Hereafter, without any confusion,
we drop the subscript ``1'' in all the 1D thermodynamic variables.

In contrast, the universal function for superfluid density $f_{s}^{3D}\left(t\right)$
remains elusive \cite{Taylor2008,Salasnich2010,Baym2013}. In this
work, we will use a phenomenological ansatz for the three-dimensional
superfluid fraction $(n_{s}/n)^{3D}=f(T/T_{c})$, following the strategy
used in Ref. \cite{Hou2013}. Thus, recalling that $T/T_{c}=[f_{n}^{3D}(t)/f_{n}^{3D}(t_{c}\simeq2.49)]^{-2/3}$,
the universal function $f_{s}^{3D}\left(t\right)$ is given by,
\begin{equation}
f_{s}^{3D}\left(t\right)=f_{n}^{3D}\left(t\right)f\left\{ \left[\frac{f_{n}^{3D}\left(t\right)}{f_{n}^{3D}\left(t_{c}\simeq2.49\right)}\right]^{-2/3}\right\} .\label{eq:fs3d}
\end{equation}
In the following, we will use the phenomenological superfluid fraction
\cite{Hou2013} 
\begin{equation}
f\left(\frac{T}{T_{c}}\right)=1-\left(\frac{T}{T_{c}}\right)^{4},\label{eq:fttc}
\end{equation}
unless otherwise stated.

\section{1D simplified two-fluid hydrodynamic equations}

Using 1D thermodynamic variables in the standard Landau two-fluid
hydrodynamic description \cite{Landau1941}, it is straightforward
to write down the simplified 1D two-fluid hydrodynamic equations.
As discussed in the previous work \cite{Taylor2005,Taylor2008,Taylor2009,Hou2013},
the solutions of these equations with frequency $\omega$ at temperature
$T$ can be derived by minimizing a variational action, which, in
terms of displacement fields $u_{s}(z)$ and $u_{n}(z)$, is given
by,
\begin{eqnarray}
S^{(2)} & = & \frac{1}{2}\int dz\left[m\omega^{2}\left(n_{s}u_{s}^{2}+n_{n}u_{n}^{2}\right)-\left(\frac{\partial\mu}{\partial n}\right)_{s}\left(\delta n\right)^{2}\right.\nonumber \\
 &  & \left.-2\left(\frac{\partial T}{\partial n}\right)_{s}\delta n\delta s-\left(\frac{\partial T}{\partial s}\right)_{n}\left(\delta s\right)^{2}\right].\label{eq:usunAction}
\end{eqnarray}
Here, $n_{s}(z)$ and $n_{n}(z)=n(z)-n_{s}(z)$ are the reduced 1D
superfluid and normal-fluid densities. $\delta n(z)\equiv-\partial(n_{s}u_{s}+n_{n}u_{n})/\partial z$
and $\delta s(z)\equiv-\partial(su{}_{n})/\partial z$ are the density
and entropy fluctuations, respectively. The effect of the weak axial
trapping potential $V_{ext}(z)=m\omega_{z}^{2}z^{2}/2$ enters Eq.
(\ref{eq:usunAction}) via the position dependence of the equilibrium
thermodynamic variables, within the local density approximation.

In superfluid helium, the solutions of the hydrodynamic action Eq.
(\ref{eq:usunAction}) can be well classified as density and temperature
waves, which are the pure in-phase mode with $u_{s}=u_{n}$ and the
pure out-of-phase mode with $n_{s}u_{s}+n_{n}u_{n}=0$, referred to
as first and second sound, respectively \cite{Landau1941}. We may
use the similar characterization for a unitary Fermi gas. To this
aim, it is useful to rewrite the action Eq. (\ref{eq:usunAction})
in terms of the displacement fields $u_{a}=(n_{s}u_{s}+n_{n}u_{n})/n$
and $u_{e}=u_{s}-u_{n}$, since the density and temperature fluctuations
can be expressed by $\delta n=-\partial(nu_{a})/\partial z$ and $\delta T=(\partial T/\partial s)_{n}\partial(sn_{s}u_{e}/n)/\partial z$,
respectively. Making use of standard thermodynamic identities, we
find that 
\begin{equation}
S^{(2)}=\frac{1}{2}\int dz\left[\mathcal{S}^{(a)}+\mathcal{S}^{(e)}+\mathcal{S}^{(ae)}\right],\label{eq:uaueAction}
\end{equation}
 where
\begin{align}
\mathcal{S}^{(a)} & =m\left(\omega^{2}-\omega_{z}^{2}\right)nu_{a}^{2}-n\left(\frac{\partial P}{\partial n}\right)_{\bar{s}}\left(\frac{\partial u_{a}}{\partial z}\right)^{2},\\
\mathcal{S}^{(e)} & =m\omega^{2}\frac{n_{s}n_{n}}{n}u_{e}^{2}-\left(\frac{\partial T}{\partial s}\right)_{n}\left[\frac{\partial}{\partial z}\left(\frac{sn_{s}}{n}u_{e}\right)\right]^{2},\\
\mathcal{S}^{(ae)} & =2\left(\frac{\partial P}{\partial s}\right)_{n}\frac{\partial u_{a}}{\partial z}\frac{\partial}{\partial z}\left(\frac{sn_{s}}{n}u_{e}\right).
\end{align}
In the absence of the coupling term $\mathcal{S}^{(ae)}$, it is clear
that the first sound mode, describe by $\mathcal{S}^{(a)}$, is the
exact solution for pure density oscillations (i.e., $u_{e}=0$ or
$\delta T=0$), while the second sound mode given by $\mathcal{S}^{(e)}$
corresponds to pure temperature oscillations with $u_{a}=0$ or $\delta n=0$.
For a uniform superfluid ($V_{ext}=0$), the solutions of $\mathcal{S}^{(a)}$
and $\mathcal{S}^{(e)}$ are plane waves of wave vector $q$ with
dispersion $\omega_{1}=c_{1}q$ and $\omega_{2}=c_{2}q$, where $mc_{1}^{2}=(\partial P/\partial n)_{\bar{s}}$
and 
\begin{equation}
mc_{2}^{2}=k_{B}T\frac{\bar{s}^{2}}{\bar{c}_{v}}\frac{n_{s}}{n_{n}}.\label{eq:c2_cv}
\end{equation}
These first and second sound velocities are the standard results used
to describe superfluid helium \cite{Landau1941}.

\begin{figure}
\begin{centering}
\includegraphics[clip,width=0.48\textwidth]{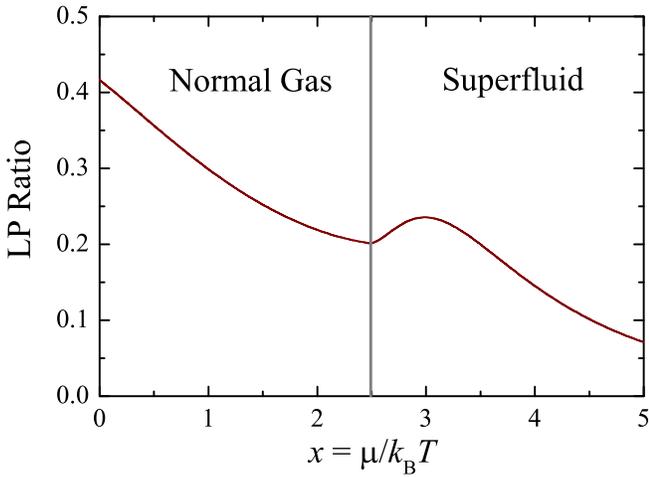} 
\par\end{centering}

\caption{(Color online) Landau-Placzek parameter $\epsilon_{\textrm{LP}}$
as a function of $x=\mu/(k_{B}T)$ in a highly elongated unitary Fermi
gas.}

\label{fig2} 
\end{figure}

In general, the coupling term $\mathcal{S}^{(ae)}$ is nonzero. Actually,
in our case, as $(\partial P/\partial s)_{n}=2T/5$, the first and
second sound are necessarily coupled at any finite temperature. This
coupling can be conveniently characterized by the dimensionless Landau-Placzek
(LP) parameter $\epsilon_{\textrm{LP}}\equiv\gamma-1$ \cite{Hu2010},
where $\gamma\equiv\bar{c}_{p}/\bar{c}_{v}$ is the ratio between
the specific heats per particle at constant pressure and density.
Indeed, with the coupling term, the second sound velocity $c_{2}$
may be well approximated by \cite{Hu2010,Hou2013}
\begin{equation}
mc_{2}^{2}=k_{B}T\frac{\bar{s}^{2}}{\bar{c}_{p}}\frac{n_{s}}{n_{n}},\label{eq:c2_cp}
\end{equation}
which differs with Eq. (\ref{eq:c2_cv}) by a factor of $\gamma=\bar{c}_{p}/\bar{c}_{v}$.
Thus, the LP ratio is a useful parameter to estimate the coupling
between first and second sound. 

In superfluid helium, $\bar{c}_{p}\simeq\bar{c}_{v}$ or $\epsilon_{\textrm{LP}}\simeq0$,
indicating that the first and second sound are well decoupled. For
a unitary Fermi gas in highly elongated traps, we have calculated
the LP ratio using the 1D thermodynamic data. As shown in Fig. \ref{fig2},
the ratio is less than $1/4$ in the whole superfluid phase. Therefore,
similar to superfluid liquid helium, the solutions of two-fluid hydrodynamic
equations for a highly elongated unitary Fermi gas are well approximated
as weakly coupled first and second sound modes. 

We note that, in the presence of axial harmonic traps ($V_{ext}\neq0$),
the actions $\mathcal{S}^{(a)}$ and $\mathcal{S}^{(e)}$ have been
solved analytically by Hou, Pitaevskii and Stringari using simple
variational ansatz \cite{Hou2013}. The coupling between first and
second sound due to $\mathcal{S}^{(ae)}$ has also been briefly commented.
In this work, by presenting the full variational calculations, we
will show how the low-lying second sound modes are affected by the
coupling. In particular, we focus on the density fluctuations of second
sound modes, which are the key observable in real experiments \cite{Tey2013}.

\subsection{Variational approach}

We assume the following polynomial ansatz for the displacement fields:
\begin{align}
u_{a}(z) & =\sum_{i=0}^{N_{p}-1}A_{i}\tilde{z}^{i},\\
u_{e}\left(z\right) & =\sum_{i=0}^{N_{p}-1}B_{i}\tilde{z}^{i},
\end{align}
where the number of the variational parameters $\{A_{i},B_{i}\}$
is $2N_{p}$, and $\tilde{z}\equiv z/Z_{F}$ is the dimensionless
coordinate with $Z_{F}$ being the Thomas-Fermi radius along the weakly
confined axial direction. Inserting this ansatz into the action Eq.
(\ref{eq:uaueAction}), the mode frequencies are obtained by minimizing
the resulting expression with respect the $2N_{p}$ parameters. The
precision of our variational calculations can be improved by increasing
the value of $N_{p}$. 

In greater detail, it is easy to see that, the action is given by
\begin{equation}
S^{(2)}=\frac{1}{2}\Lambda^{\dagger}\mathcal{S}\left(\omega\right)\Lambda,
\end{equation}
where $\Lambda\equiv\left[A_{0},B_{0},\cdots,A_{i},B_{i},\cdots,A_{N_{p}-1},B_{N_{p}-1}\right]^{T}$
and $\mathcal{S}(\omega)$ is a $2N_{p}\times2N_{p}$ matrix with
block elements,
\begin{equation}
\left[\mathcal{S}\left(\omega\right)\right]_{ij}\equiv\left[\begin{array}{cc}
M_{ij}^{(a)}\omega^{2}-K_{ij}^{(a)} & -K_{ij}^{(ae)}\\
-K_{ji}^{(ae)} & M_{ij}^{(e)}\omega^{2}-K_{ij}^{(e)}
\end{array}\right].\label{eq:SW}
\end{equation}
Here, we have introduced the weighted mass moments,
\begin{align}
M_{ij}^{(a)} & =m\int dz\tilde{z}^{i+j}n\left(z\right),\\
M_{ij}^{(e)} & =m\int dz\tilde{z}^{i+j}\left[\frac{n_{s}n_{n}}{n}\right]\left(z\right),
\end{align}
and the spring constants,
\begin{align}
K_{ij}^{(a)} & =\frac{7}{5}ij\int dz\tilde{z}^{i+j}P\left(z\right)/z^{2}+\omega_{z}^{2}M_{ij}^{(a)},\\
K_{ij}^{(ae)} & =\frac{2T}{5}i\left(i-1\right)\int dz\tilde{z}^{i+j}z^{-2}\left[\frac{sn_{s}}{n}\right]\left(z\right),\\
K_{ij}^{(e)} & =\int dz\left(\frac{\partial T}{\partial s}\right)_{n}\frac{\partial}{\partial z}\left(\frac{sn_{s}\tilde{z}^{i}}{n}\right)\frac{\partial}{\partial z}\left(\frac{sn_{s}\tilde{z}^{j}}{n}\right).
\end{align}
In deriving $K_{ij}^{(a)}$ and $K_{ij}^{(ae)}$, we have used the
universal relations satisfied by the highly elongated unitary Fermi
gas: $n(\partial P/\partial n)_{\bar{s}}=7P/5$ and $(\partial P/\partial s)_{n}=2T/5$.
For a given value of $\mu/k_{B}T$ (or $T/T_{F}$, see below), the
weighted mass moments and spring constants can be calculated by using
local thermodynamic variables in Eqs. (\ref{eq:pressure1d}), (\ref{eq:density1d}),
(\ref{eq:entropy1d}), (\ref{eq:cv1d}) and (\ref{eq:sfdensity1d}).
We note that the universal scaling function for superfluid density
is given by Eqs. (\ref{eq:fs}), (\ref{eq:fs3d}) and (\ref{eq:fttc}). 

In practice, the minimization of the action $S^{(2)}$ is equivalent
to solving 
\begin{equation}
\mathcal{S}\left(\omega\right)\Lambda=0.
\end{equation}
 Once a solution (i.e., the $k$-th mode frequency $\omega_{k}$ and
the coefficient eigenvector $\Lambda_{k}$) is found, we calculate
the density fluctuation of the mode, by using
\begin{equation}
\delta n_{k}\left(z\right)=-\sum_{i=0}^{N_{p}-1}A_{i}^{(k)}\frac{\partial}{\partial z}\left[n\left(z\right)\tilde{z}^{i}\right].\label{eq:dstyfluct}
\end{equation}

\section{Results and discussions}

We have performed numerical calculations for the number of the variational
parameter $N_{p}$ up to $24$, for any given temperature $T/T_{F}$
or chemical potential $\mu/k_{B}T$. By recalling that the Fermi temperature
$T_{F}$ of a three-dimensional trapped Fermi gas is given by
\begin{equation}
k_{B}T_{F}=\hbar\left(3N\omega_{\perp}^{2}\omega_{z}\right)^{1/3}
\end{equation}
and the number of atoms $N=\int dzn(z)$, these two parameters are
related by,
\begin{equation}
\frac{T}{T_{F}}=\left[\frac{3}{\sqrt{\pi}}\int_{0}^{\infty}dt\frac{1}{\sqrt{t}}f_{n}(\frac{\mu}{k_{B}T}-t)\right]^{-1/3}.
\end{equation}
In the following, we first discuss the decoupled first and second
sound, in connection with the previous results by Hou, Pitaevskii
and Stringari \cite{Hou2013}. Then, we focus on the effect of the
mode coupling.

\subsection{Decoupled first sound}

In Ref. \cite{Hou2013}, the action $\mathcal{S}^{(a)}$ has been
solved by using the ansatz $u_{a}^{k=2}(z)=A_{2}z^{2}+A_{0}$ and
$u_{a}^{k=3}(z)=A_{3}z^{3}+A_{1}z$ for the $k=2$ and $k=3$ first
sound modes, respectively. Here, $k$ is the index of a mode and counts
the number of nodes ($=k+1$) in its density fluctuation. These are
the first two solutions, whose frequency varies with increasing temperature,
due to the non-trivial temperature dependence of equation of state
\cite{Hou2013}. Indeed, it is easy to prove that 
\begin{equation}
K_{ij}^{(a)}=\left[\frac{7}{5}\frac{ij}{\left(i+j-1\right)}+1\right]\omega_{z}^{2}M_{ij}^{(a)}.
\end{equation}
Thus, if $i=0$ or $j=0$, we have $K_{ij}^{(a)}=\omega_{z}^{2}M_{ij}^{(a)}$.
Together with the fact that $K_{i=0,j}^{(ae)}=0$ or $K_{j=0,i}^{(ae)}=0$,
it is clear that the $k=0$ first sound mode with variational ansatz
$u_{a}^{k=0}(z)=A_{0}$ is an exact solution of the two-fluid hydrodynamic
equations. In fact, it is precisely the undamped \emph{dipole} oscillation,
with invariant frequency $\omega=\omega_{z}$. Similarly, in the case
of $i=1$ or $j=1$, $K_{ij}^{(a)}=(12/5)\omega_{z}^{2}M_{ij}^{(a)}$,
revealing that the $k=1$ first sound mode - the \emph{breathing}
mode - is another exact solution of the two-fluid hydrodynamic equation
with invariant frequency $\omega=\sqrt{12/5}\omega_{z}$ \cite{Hou2013,Hou2013b}.

\begin{figure}
\begin{centering}
\includegraphics[clip,width=0.48\textwidth]{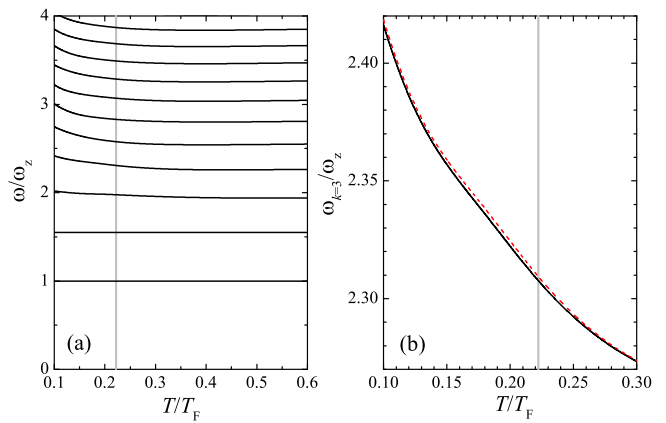} 
\par\end{centering}

\caption{(Color online) (a) Temperature dependence of the first sound mode
frequencies. (b) Enlarged view for the fourth first sound mode frequency.
The red dashed line shows the result with the ansatz $u_{a}(z)=A_{3}z^{3}+A_{1}z$,
obtained earlier by Hou, Pitaevskii and Stringari \cite{Hou2013}.
The vertical grey lines show the critical temperature, $T_{c}\simeq0.223T_{F}$.}

\label{fig3} 
\end{figure}

In Fig. \ref{fig3}(a), we report the variational results for first
sound mode frequencies with $N_{p}=24$. In agreement with the observation
by Hou, Pitaevskii and Stringari \cite{Hou2013}, we find that $u_{a}^{k=2}(z)$
and $u_{a}^{k=3}(z)$ provide excellent variational ansatz for the
third and fourth modes. As shown in Fig. \ref{fig3}(b), the higher-order
correction, for example, for the $k=3$ mode, is at the order of $0.1\%$
in relative and can only be seen in the vicinity of the critical temperature.

\subsection{Decoupled second sound}

\begin{figure}
\begin{centering}
\includegraphics[clip,width=0.48\textwidth]{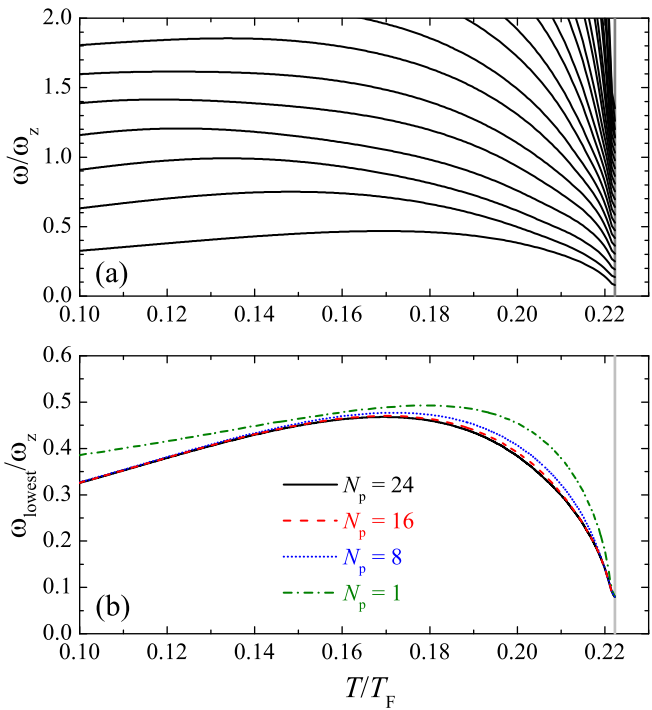} 
\par\end{centering}

\caption{(Color online) (a) Temperature dependence of the second sound mode
frequencies with $N_{p}=24$. The mode frequencies vanish right at
superfluid phase transition. However, our numerical calculations become
less accurate slightly below the transition and can not produce correctly
the zero frequency exactly at $T_{c}$. (b) Enlarged view for the
lowest second sound mode frequency. The mode frequency converges with
increasing the number of variational ansatz $N_{p}$. The result with
$N_{p}=1$ (green dot-dashed line) corresponds to a constant displacement
field $u_{e}$ \cite{Hou2013}. The vertical grey lines show the critical
temperature.}

\label{fig4} 
\end{figure}

In Fig. \ref{fig4}(a), we present the results for the mode frequency
of decoupled second sounds. With increasing temperature, the mode
frequency initially increases and reaches a maximum before finally
dropping to zero at the superfluid phase transition \cite{Hou2013}.
In sharp contrast to first sound modes, the convergence of the polynomial
ansatz for second sound modes appears to be slow. For the lowest dipole
second sound mode, our variational approach only converges at $N_{p}\geq16$,
as can be seen from Fig. \ref{fig4}(b). Moreover, compared with the
fully converged result, a constant displacement field $u_{e}$ (i.e.,
$N_{p}=1$) can lead to a relative error as large as $20\%$ close
to the superfluid phase transition. For the higher order second sound
modes, we observe that the convergence of the polynomial ansatz becomes
even slower.

\subsection{Full solutions of 1D two-fluid hydrodynamics}

\begin{figure}
\begin{centering}
\includegraphics[clip,width=0.48\textwidth]{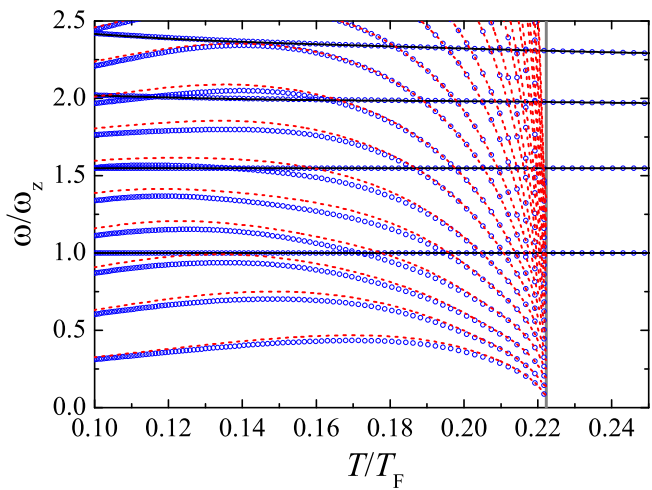} 
\par\end{centering}

\caption{(Color online) Temperature dependence of the full two-fluid hydrodynamic
mode frequencies (blue circles). For comparison, we show also the
mode frequencies of the decoupled first and second sounds, respectively,
by black solid and red dashed lines. The vertical grey line indicates
the critical temperature.}

\label{fig5} 
\end{figure}

We now include the mode coupling term $\mathcal{S}^{(ae)}$. In Fig.
\ref{fig5}, we report the full variational results with $N_{p}=24$
by blue circles. For comparison, we show also the decoupled first
and second sound mode frequencies, by using black lines and red dashed
lines. As anticipated, at the qualitative level, the full variational
predictions can be well approximated by the decoupled results, confirming
our previous idea that in highly elongated harmonic traps, the solutions
of two-fluid hydrodynamics of a unitary Fermi gas can indeed be viewed
as weakly coupled first and second sound modes.

\begin{figure}
\begin{centering}
\includegraphics[clip,width=0.48\textwidth]{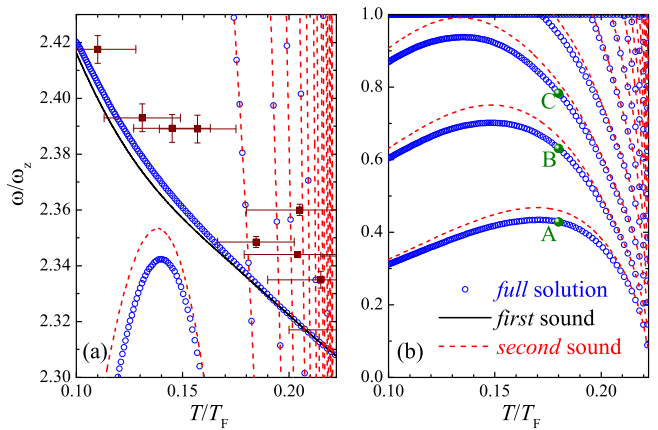} 
\par\end{centering}

\caption{(Color online) Blow-up of the full two-fluid hydrodynamic mode frequencies
(blue circles), near the $k=3$ first sound mode (a) and the lowest
second sound modes (b). In (a), the experimental data from the Innsbruck
experiment \cite{Tey2013} are shown by brown solid squares with error
bars.}

\label{fig6} 
\end{figure}

At the \emph{quantitative} level, the first sound mode is nearly unaffected
by the coupling term $\mathcal{S}^{(ae)}$. This is evident in Fig.
\ref{fig6}(a), where we present an enlarged view for the $k=3$ first
sound mode. The mode frequency has been pushed up by about $0.5\%$
at $T\sim0.15T_{F}$ by the coupling. Experimentally, the frequency
of the $k=3$ first sound mode has been measured very recently \cite{Tey2013,Guajardo2013}.
In the figure, we show the experimental data by solid squares with
error bars. It is known that the data systematically lie above the
variational result with $u_{a}^{k=3}(z)$ \cite{Tey2013,Guajardo2013,Hou2013}.
Our full variational predictions seem to agree better with the experimental
data. However, the improvement is too slight to account for the discrepancy.

On the other hand, the frequency of second sound modes is notably
pushed down by the coupling, as shown in Fig. \ref{fig6}(b). The
maximum correction is up to $10\%$ when the temperature is about
$0.15T_{F}$. Therefore, for a quantitative prediction of second sound
modes over the whole temperature regime, we must fully solve the coupled
Landau two-fluid hydrodynamic equations.

\begin{figure}
\begin{centering}
\includegraphics[clip,width=0.48\textwidth]{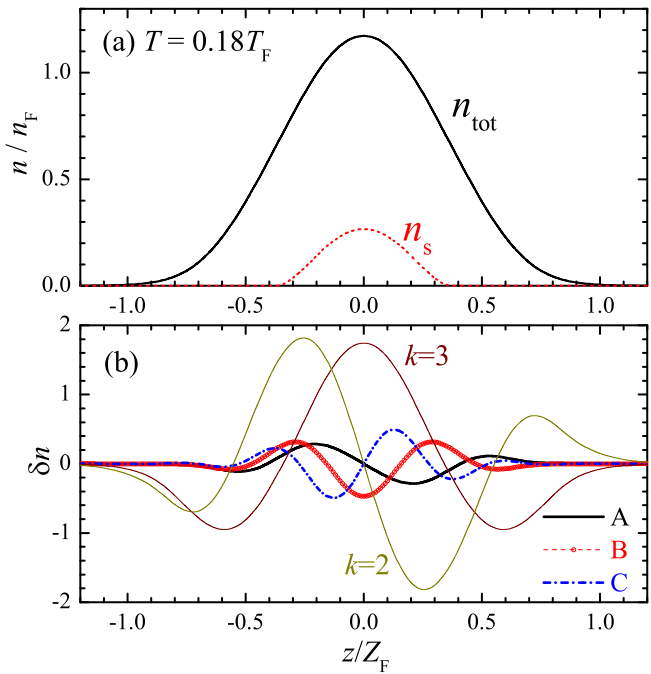} 
\par\end{centering}

\caption{(Color online) (a) Density distribution (solid line) and superfluid
density distribution (dashed line) at $T=0.18T_{F}$, in units of
the peak linear density of an ideal Fermi gas at the trap center ($n_{F}$).
(b) Density fluctuations of the $k=2$ and $k=3$ first sound modes
(thin lines) and of the three lowest second sound modes (thick lines),
at the frequencies indicated in Fig. \ref{fig6}(b) by A, B and C.
The amplitude of the second-sound density fluctuations is about $1/3$
of that of the first-sound density fluctuations. }

\label{fig7} 
\end{figure}

\subsection{Density measurement of discretized second sound modes}

The sizable correction in mode frequencies strongly indicates that,
the density fluctuation $\delta n_{k}(z)$ of a second sound mode,
as a result of its coupling to first sound modes, could also be significant.
In Fig. \ref{fig7}(b), we show the density fluctuations of the lowest
three second sound modes at the temperature $T=0.18T_{F}$ (thick
lines), in relative to the density fluctuations of the $k=2$ and
$k=3$ first sound modes (thin lines). The absolute amplitude of density
fluctuations depends on the detailed excitation scheme used in experiments.

In a recent investigation by the Innsbruck group \cite{Tey2013,Guajardo2013},
the density fluctuations of the $k=2$ and $k=3$ first sound modes
have been excited and measured, to a reasonable accuracy. For the
detailed resonant excitation scheme, we refer to the experimental
papers of Refs. \cite{Tey2013} and \cite{Guajardo2013}. We anticipate
that the similar excitation procedure works also for second sound
modes, by carefully choosing the position and size of excitation laser
beam, which provide best mode matching. As shown in Fig. \ref{fig7}(b),
it is remarkable that the amplitude of the second sound density fluctuations
is at the same order as that of the $k=2$ and $k=3$ first sound
modes, over a useful range of temperatures. As we shall see below
from the analysis of the density response function, this implies that
the low-lying second sound mode could be observed by looking at its
density fluctuation, after a proper excitation.

To obtain the density response function, we consider adding a density
perturbation of the form $\delta V(z,t)=\lambda f(z)e^{-i\Omega t}$
to the equilibrium two-fluid equations, where $\lambda$ is the strength
of the perturbation and $f(z)$ is a normalized shape function (i.e.,
$\int dzf^{2}(z)=1$). This leads to a density fluctuation $\delta n(z,t)$
with its amplitude proportional to $\lambda$ and in turn gives the
response function $\chi_{nn}(\Omega;f)$ defined by 
\begin{equation}
\lambda\chi_{nn}\left(\Omega;f\right)e^{-i\Omega t}=\int dzf\left(z\right)\delta n\left(z,t\right).
\end{equation}
Note that in general the response function depends on the form of
the perturbation $f(z)$. In greater detail, the density perturbation
generates an additional term $\delta S^{(2)}$ in the two-fluid hydrodynamic
action,
\begin{equation}
S^{\left(2\right)}=\frac{1}{2}\Lambda_{\Omega}^{\dagger}\mathcal{S}\left(\Omega\right)\Lambda_{\Omega}+\delta S^{(2)},
\end{equation}
where
\begin{equation}
\delta S^{(2)}=\lambda\sum_{i=0}^{N_{p}-1}\int dzf\left(z\right)\frac{\partial}{\partial z}\left[n\left(z\right)\tilde{z}^{i}\right]A_{i}^{(\Omega)}.
\end{equation}
Here we have used the index ``$\Omega$'' to distinguish $\Lambda_{\Omega}$
from the eigenvector $\Lambda$ obtained by diagonalizing the action
matrix Eq. (\ref{eq:SW}). By introducing a vector $F=[f_{0},0,f_{1},0,\cdots,f_{N_{p}-1},0]^{T}$,
where $f_{i}\equiv\int dzf(z)\partial[n\left(z\right)\tilde{z}^{i}]/\partial z$,
we obtain 
\begin{equation}
\Lambda_{\Omega}=-\lambda\mathcal{S}^{-1}\left(\Omega\right)F
\end{equation}
 by minimizing the perturbed action. Substituting this result into
the expression for the density fluctuation Eq. (\ref{eq:dstyfluct}),
we find that,
\begin{align}
\chi_{nn}\left(\Omega;f\right) & =F^{\dagger}\mathcal{S}^{-1}(\Omega)F,\nonumber \\
 & =\sum_{k}\frac{Z_{k}}{2\omega_{k}}\left(\frac{1}{\Omega-\omega_{k}}-\frac{1}{\Omega+\omega_{k}}\right),
\end{align}
where 
\begin{equation}
Z_{k}=\left|F^{\dagger}\Lambda_{k}\right|^{2}=\left|\int dzf\left(z\right)\delta n_{k}\left(z\right)\right|^{2}\label{eq:Bk}
\end{equation}
 is the \emph{residue} of the response function for the $k$-th collective
mode with eigenvector $\Lambda_{k}$ and frequency $\omega_{k}$ \cite{note}.

Using Eq. (\ref{eq:Bk}), it is clear that a sizable density fluctuation
$\delta n_{k}\left(z\right)$ of second sound modes - as reported
in Fig. \ref{fig7} - implies that a significant residue in the density
response for second sound can be achieved by optimizing the \emph{unit}
shape function $f\left(z\right)$ such that $f\left(z\right)\propto\delta n_{k}\left(z\right)$.
Therefore, discrete second sound could be excited in a similar way
as first sound, without imposing a large strength $\lambda$ for the
density perturbation $\delta V(z,t)$.

\subsection{Dependence on the superfluid fraction}

\begin{figure}
\begin{centering}
\includegraphics[clip,width=0.48\textwidth]{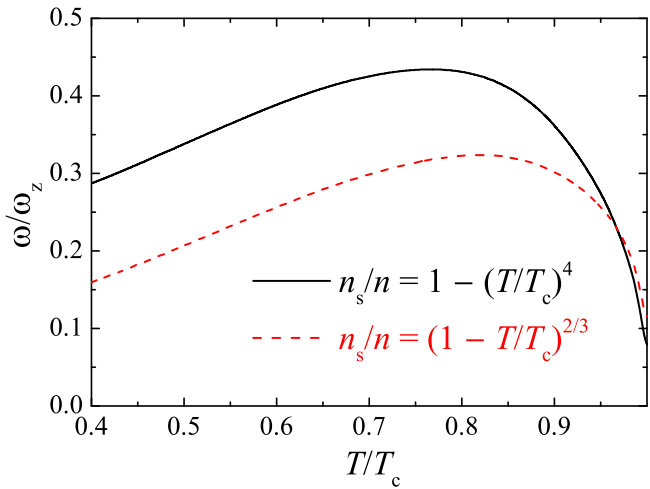} 
\par\end{centering}

\caption{(Color online) Sensitivity of the lowest second sound mode on the
superfluid density. The critical temperature $T_{c}\simeq0.223T_{F}$.}

\label{fig8} 
\end{figure}

We so far restrict ourselves to the phenomenological superfluid fraction
Eq. (\ref{eq:fttc}). In Fig. \ref{fig8}, we report the dependence
of the lowest second sound frequency on the form of superfluid fraction.
The sensitive dependence indicates that practically the unknown superfluid
density of a unitary Fermi gas could be accurately determined by measuring
the mode frequency of low-lying second sound modes.

\section{Conclusions}

In conclusion, using a variational approach, we have fully solved
the one-dimensional simplified Landau two-fluid hydrodynamic equations,
which describe the collective excitations of a unitary Fermi gas in
highly elongated harmonic traps. Resembling the superfluid helium,
the solutions are well characterized by weakly coupled first and second
sound modes. Discretized first and second sound mode frequencies have
been accurately predicted. 

Though the coupling between first and second sound is weak, it still
induces significant density fluctuations for second sound modes, suggesting
that second sound could be observed by measuring the density fluctuations
after properly modulating the axial harmonic trapping potential. Owing
to the very high precision in the frequency calibration, the experimental
measurement of discretized second sound mode frequency provides a
promising way to accurately determining the superfluid density of
a unitary Fermi gas, which remains elusive to date.

Ideally, we anticipate that the relative error in the measurement
of the second sound mode frequency is about $0.5\%$. For the low-lying
modes, whose mode frequency is smaller than $\omega_{z}$, the damping
rate might be reasonably small. By assuming a superfluid fraction
in the form,
\begin{equation}
f\left(\frac{T}{T_{c}}\right)=\left(1-\frac{T}{T_{c}}\right)^{2/3}\left[a_{0}+a_{1}\left(\frac{T}{T_{c}}\right)+\cdots\right],\label{eq:fttc_fitting}
\end{equation}
which correctly reproduces the critical behavior near superfluid phase
transition, we may determine the parameters $\{a_{0},a_{1},a_{2},\cdots\}$
by fitting the experimental data to the full variational predictions
for the discretized low-lying second sound mode frequencies.
\begin{acknowledgments}
We acknowledge the useful discussion with Yan-Hua Hou, which stimulated
this research. We thank Meng Khoon Tey for sending us the experimental
data of the $k=2$ and $k=3$ first sound modes. Our work was supported
by the ARC Discovery Projects (Grant Nos. DP0984637, DP140100637,
DP0984522, DP140103231 and FT130100815) and NFRP-China (Grant No.
2011CB921502).\end{acknowledgments}

\end{document}